\documentclass[%
 reprint, 
 amsmath,amssymb,
 aps,
citeautoscript,
prb,
floatfix]{revtex4-1}
\usepackage{graphicx}%
\usepackage{dcolumn}%
\usepackage{bm}%
\usepackage{xcolor}
\usepackage[normalem]{ulem}
\usepackage{color,soul}
\usepackage{nccmath}
\usepackage[]{lineno}
\usepackage{rotating} 
\usepackage{booktabs} 
\usepackage{amsmath} 

\begin{document}

\preprint{APS/123-QED}

\title{Bright Pulsed Squeezed Light for Quantum-Enhanced Precision
Microscopy}

\author{Alex Terrasson\textsuperscript{1}}
\author{Lars Madsen\textsuperscript{1}}
\author{Joel Q. Grim\textsuperscript{2}}
\author{Warwick.~P. Bowen\textsuperscript{1,{$\star$}}\vspace*{3 mm}}

\affiliation{%
\textsuperscript{1} \mbox{Australian Research Council Centre of Excellence in Quantum Biotechnology, University of Queensland}
}%
\affiliation{%
\textsuperscript{2} \mbox{U.S. Naval Research Laboratory.}
}%
\affiliation{%
\textsuperscript{{$\star$} }w.bowen@uq.edu.au \vspace*{-6 mm}
}%

\pacs{Valid PACS appear here}
\maketitle

{\bf
Squeezed states of light enable enhanced measurement precision by reducing noise below the standard quantum limit. 
A key application of squeezed light is nonlinear microscopy, where state-of-the-art performance is limited by photodamage and quantum-limited noise.
Such microscopes require bright, pulsed light for optimal operation, yet generating and detecting bright pulsed squeezing at high levels remains challenging.
In this work, we present an efficient technique to generate high levels of bright picosecond pulsed squeezed light using a $\chi^2$ optical parametric amplification process in a waveguide.
We measure $-3.2~\mathrm{dB}$ of bright squeezing with optical power compatible with nonlinear microscopy, as well as $-3.6~\mathrm{dB}$ of vacuum squeezing. Corrected for losses, these squeezing levels correspond to $-15.4^{+2.7}_{-8.7}~\mathrm{dB}$ of squeezing generated in the waveguide. The measured level of bright amplitude pulsed squeezing is to our knowledge the highest reported to date, and will contribute to the broader adoption of quantum-enhanced nonlinear microscopy in biological studies.
}

\maketitle


\section{Introduction}\label{sec_intro}

Stimulated Raman scattering (SRS) imaging is a nonlinear microscopy technique widely used in biological studies for its ability to provide label-free, quantitative, and chemically specific contrast \cite{cheng2015vibrational,camp2015chemically}. SRS has been applied to study metabolic processes \cite{zhang2019spectral}, neuron membrane potentials \cite{tian2016monitoring}, antibiotic responses \cite{schiessl2019phenazine}, nerve degeneration \cite{ tian2016monitoring}, among other biological processes. Improvements in SRS performance would enhance existing studies and enable emerging applications such as high-throughput cancer screening \cite{cheng2021emerging,tan2023profiling}. 
However photodamage to biological samples limits the usable optical power and therefore the achievable signal level. At the same time, state-of-the-art SRS microscopes are fundamentally limited by the standard quantum limit set by shot noise~\cite{saar2010video,freudiger2008label}. As further gains from classical technical improvements diminish, sub-shot noise operation represents the primary pathway toward enhanced SRS sensitivity \cite{casacio2021quantum}.

Squeezed light is a nonclassical state of light in which a nonlinear interaction redistributes the uncertainty between the two field quadratures, reducing noise in one quadrature below the quantum limit at the expense of increased noise in the other. It was identified  as capable of enhancing measurement precision below the standard quantum limit
~\cite{slusher1990quantum,davidovich1996sub,lawrie2020squeezing}, and high squeezing levels have been demonstrated  for continuous wave and femtosecond pulses ~\cite{dong2008experimental,vahlbruch2016detection}. For maximum sensitivity in SRS microscopy, amplitude squeezed light must be generated at high optical power (bright squeezing) and at picosecond (ps) or femtosecond pulse duration, with most systems using ps pulses to match the coherence time of molecular vibrations. ~\cite{xu2025pushing}. In recent years, quantum-enhanced vibrational microscopes have been demonstrated for Brillouin and Raman imaging using bright pulsed squeezed light illumination \cite{casacio2021quantum,xu2022quantum,li2022quantum,gong2023super,terrasson2024fast,li2024harnessing}. These works employ either single-mode squeezed light with direct detection or twin-beam intensity-squeezed light with balanced detection, the latter incurring a $3~\mathrm{dB}$ noise penalty compared to direct detection.
In both cases, experimental challenges associated with generating and detecting high levels of bright ps pulsed squeezing has limited the level of enhancement achieved \cite{heng2025quantum,xu2025pushing}. 

The highest reported level of ps pulsed squeezing is $-5.88~\mathrm{dB}$ \cite{amari2023highly}, achieved using a single-pass optical parametric amplifier (OPA) based on a periodically poled lithium niobate (PPLN) waveguide. This result relied on advanced temporal and spatial shaping of the local oscillator (LO).
Earlier work demonstrated comparable squeezing levels of $-5.8~\mathrm{dB}$ using LO–squeezed field co-propagation through the OPA to achieve high spatial overlap \cite{kim1994quadrature}; however, the corresponding bright squeezing reported in a subsequent study was limited to $-0.8~\mathrm{dB}$ \cite{serkland1997amplitude}. 
In our previous work, we generated and measured $-1.8~\mathrm{dB}$ of bright amplitude pulsed squeezing \cite{terrasson2024fast} using a single-pass OPA based on a PPKTP crystal. In that approach, a bright field was used to seed the OPA and direct detection was performed, with a custom detector that was able to tolerate the high peak intensities. Parasitic multimode effects associated with this configuration limited the measured squeezing level.

In this work, we present a bright, pulsed squeezed-light source capable of high levels of amplitude squeezing at power levels used in state-of-the-art SRS microscopes \cite{freudiger2008label,casacio2021quantum}. 
We generate the squeezed field in a single pass OPA using a PPLN waveguide. Spatial overlap is achieved by co-propagating the LO with the squeezed field through the waveguide. The field is then displaced by the LO into a bright squeezed field with a set of wave-plates and polarizing beam splitter (PBS).
We report $-3.2~\mathrm{dB}$ of bright squeezing in direct detection at a displaced optical power of $3.2~\mathrm{mW}$, and $-3.6~\mathrm{dB}$ of squeezing measured via homodyne detection. The observed squeezing level is primarily limited by losses, in particular the quantum efficiency (QE) of the photodetectors at 0.75. After correcting for the losses, those measured squeezing levels correspond to $-15.4^{+2.7}_{-8.7}~\mathrm{dB}$ generated within the waveguide. The spatial mode overlap between squeezed field and the LO is simulated to be 99.7\%, while the temporal mode overlap is experimentally measured to be 97.7\%. 
This experimental approach opens a pathway towards higher levels of generated and detected bright squeezed light, a key requirement for the broader adoption of quantum-enhanced nonlinear microscopy in biological studies. 

\section{Experimental setup}

\begin{figure*}[ht!]
  \centering
  \includegraphics[width=\linewidth]{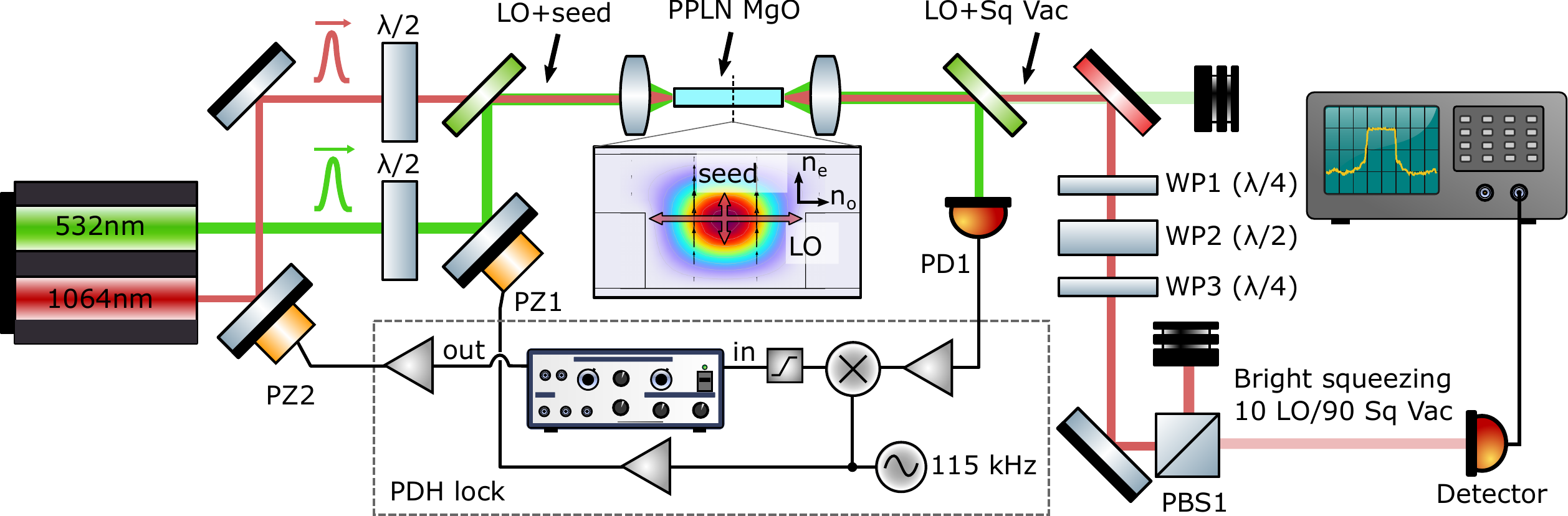} 
	\caption{\textbf{Experimental setup} A dual-head laser generates synchronized 1064nm and 532 nm light pulses with 6 ps and 80 MHz repetition rate. The beams are recombined and focused into a periodically poled MgO-doped lithium niobate (PPLN) waveguide. The 1064 beam is polarized along the ordinary axis of the waveguide and serves as the LO. A weak component polarized along the extraordinary axis seeds the OPA and generates squeezed vacuum. The 532 nm beam is detected and used to lock the relative phase between the 1064 nm and 532 nm fields using a PDH lock scheme. The LO and squeezed vaccum are recombined and mixed at a PBS using three waveplates (WP 1-3). After the PBS, the field is detected either in a balanced homodyne configuration with a 50/50 mixing ratio using two photodetectors (not shown) or with a 90/10 mixing ratio at a single output to generate bright squeezing.
 } 
	\label{fig:fig1} 
\end{figure*}

The experimental setup is shown in Figure \ref{fig:fig1}. 
A dual-head laser provides synchronized fundamental and second-harmonic beams at 1064 nm and 532 nm with 6 ps pulses and 80 MHz repetition rate. 
The beams are combined and focused by a microscope objective (PAL-20-NIR-LCOO objective, Optosigma) into a 5-mm-long periodically poled MgO-doped lithium niobate ridge waveguide with AR-coated facets(HC Photonics). 
The short waveguide length limits the effect of group velocity mismatch between the green pump and the 1064 nm seed. 
An oven (TC-038D) sets the temperature to the phase-matching temperature.

The polarization of the 1064 nm beam is adjusted such that only a small fraction seeds the OPA process. The 532 nm field pumps the OPA, generating thr squeezed vacuum. 
The rest of the 1064 nm power is polarized along the ordinary axis of the waveguide and serves as the LO. 
Single-mode confinement in the waveguide ensures high spatial mode matching between the squeezed field and the LO. 
The output fields are collected with a lens (C330TMD, NA=0.7). The second harmonic beam is separated and used as input in a Pound Drever Hall (PDH) locking scheme \cite{Black2001,casacio2021quantum} to stabilize the relative phase between the OPA pump and seed.
A set of waveplates (WP1-3) controls the relative phase between LO and squeezed vacuum and sets the splitting ratio at a polarizing beam splitter (PBS1). 
In more detail, WP1 converts the two linear polarizations into left- and right-handed circular polarizations. 
WP2 controls the relative phase between these components, thereby setting the squeezing angle. 
WP3 converts the field  back towards linear polarization, determining the splitting ratio at PBS1. 

Adjusting WP2 to amplitude squeezing and WP3 to a 90/10 splitting ratio generates a bright amplitude-squeezed beam with 10\% of the LO power (at the cost of a 10\% loss on the squeezed field). Separate phase locking of the LO and squeezed field is not required as the LO and seed co-propagate.
In this configuration, bright squeezing is measured in direct detection with a custom photodetector compatible with SRS microscopy (see previous work \cite{casacio2021quantum} for details).
Homodyne detection~\cite{schumaker1984noise} can also be performed when WP3 sets a 50/50 splitting ratio and both output ports are detected, the enables to measure quadrature squeezing in addition to amplitude squeezing. 
In both configurations, the noise spectrum is recorded using a low-noise spectrum analyzer (N9010A, Agilent).

In previous work \cite{casacio2021quantum,terrasson2024fast}, bright amplitude squeezing was generated by seeding the OPA with a bright field, eliminating the need for a spatially and temporally matched LO and compatible with direct detection. This approach, however, introduces an equivalent mode-matching requirement between the seed and the OPA pump: any seed component not overlapping with the pump is detected at the shot-noise level without deamplification, causing its relative contribution to increase with deamplification. 
To mitigate this limitation, we adopt a weakly seeded OPA followed by displacement after the OPA using a matched LO. In this configuration, imperfect LO–signal overlap reintroduces shot noise, but its contribution is not amplified, resulting in improved robustness of the squeezing measurement.

\begin{figure}[ht!]
  \centering
  \includegraphics[width=0.8\linewidth]{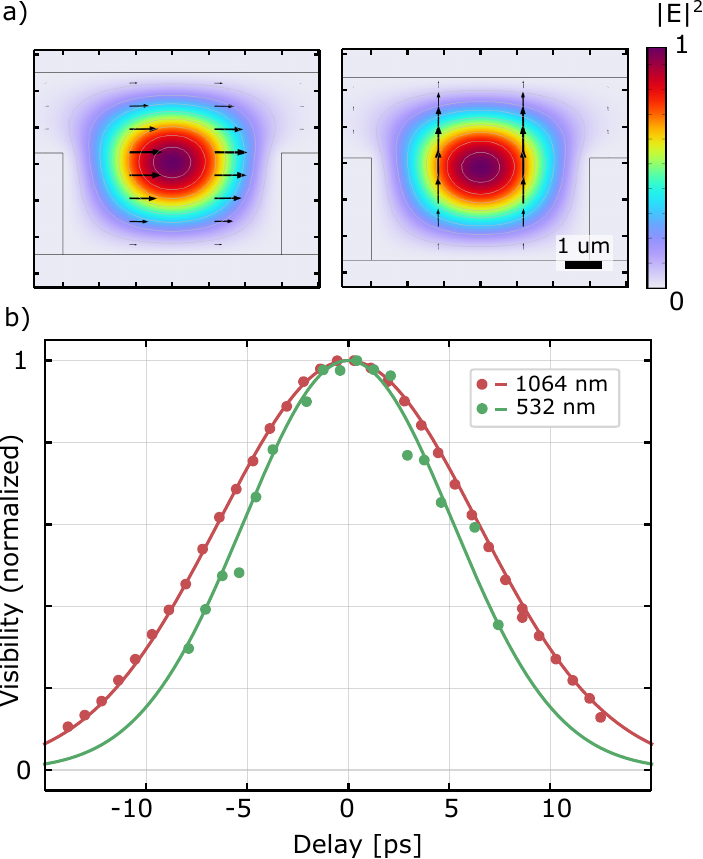} 
	\caption{\textbf{Spatial and temporal overlap between the LO and squeezed field} \textbf{a)} COMSOL modal analysis of the waveguide modes for S and P polarizations. The top panel shows the mode intensity for the horizontally polarized mode, corresponding to the LO. The bottom panel shows the mode intensity for the vertically polarized mode, corresponding to the squeezed field. \textbf{b)} Experimental measurement of the pulse durations of the LO and OPA pump. Pulse durations are measured using an interferometer by monitoring the interference visibility as the optical path length of one arm is varied. Red markers show the LO data and the solid curve is a Gaussian fit, yielding a pulse duration of $\tau_{\mathrm{LO,exp}} = 6.4 \pm 0.07~\mathrm{ps}$. Green markers correspond to the OPA pump, with a fitted pulse duration of $\tau_{\mathrm{pump,exp}} = 5.17 \pm 0.15~\mathrm{ps}$.} 
	\label{fig:fig2} 
\end{figure}

 We quantify the spatial overlap $\eta_{\mathrm{spatial}}$ between the self-matched LO and the squeezed field using finite-element simulations performed in COMSOL. The spatial overlap in the far field, where interference occurs at PBS1, is identical to the modal overlap in the waveguide. We therefore simulate the guided modes along the ordinary and extraordinary waveguide axes, with refractive indices of 2.2288 and 2.1474, respectively \cite{zelmon1997infrared}. 
The corresponding 2D spatial mode profiles are shown in Figure \ref{fig:fig2}, a). The optical confinement provided by the waveguide yields a spatial overlap of $\eta_{\mathrm{spatial}} = 0.997$.

In an ideal OPA process, the squeezed field has the same pulse duration as the pump; therefore, we experimentally characterize temporal overlap by measuring the LO and pump pulse durations. This is done using a free space interferometric autocorrelation setup where the interference fringes are scanned with a piezo-mounted mirror, and the visibility is recorded as a function of delay.
The results are shown Figure \ref{fig:fig2}, b) where the red and green markers corresponds to the measured visibilities for the 1064 and 532 nm, respectively, and the solid lines are a Gaussian fit to the data. From these fits, we extract pulse durations of $\tau_{\mathrm{LO,exp}} = 6.4 \pm 0.07~\mathrm{ps}$ and $\tau_{\mathrm{pump,exp}} = 5.17 \pm 0.15~\mathrm{ps}$, yielding an experimental temporal overlap of $\eta_{\mathrm{temp,exp}} = 0.977 \pm 0.006$. The measured pump pulse duration is 14\% longer than the ideal value, which we attribute to group velocity broadening in the second-harmonic generation stage. 

Temporally shaping the LO to match the squeezed field pulse duration can further increase the temporal overlap. This approach has been demonstrated by propagating the LO through a second optical parametric amplifier (OPA) \cite{eto2011efficient}. In that work, however, imperfect temporal overlap accounted for a significant fraction of the total optical loss. In the present system, the measured temporal overlap results in only 2.3\% loss. Moreover, temporal shaping of the LO using an OPA modifies its photon-number statistics. While this does not affect homodyne detection due to common-mode rejection in the balanced measurement, it can degrade squeezing measurements based on direct detection, where the photon statistics of the LO directly contribute to the detected noise.

\section{Squeezed light measurement}\label{secSHG}

We measure squeezing using both balanced homodyne detection and direct detection. Pulsed quadrature squeezing is characterized using balanced homodyne detection. The homodyne detector is balanced by adjusting WP3 to achieve a 50/50 splitting ratio at PBS1, with $3.2~\mathrm{mW}$ of LO power incident on each photodetector. The vacuum squeezing level is obtained by subtracting the photocurrents from the two detector outputs.

\begin{figure}[ht!]
  \centering
  \includegraphics[width=0.9\linewidth]{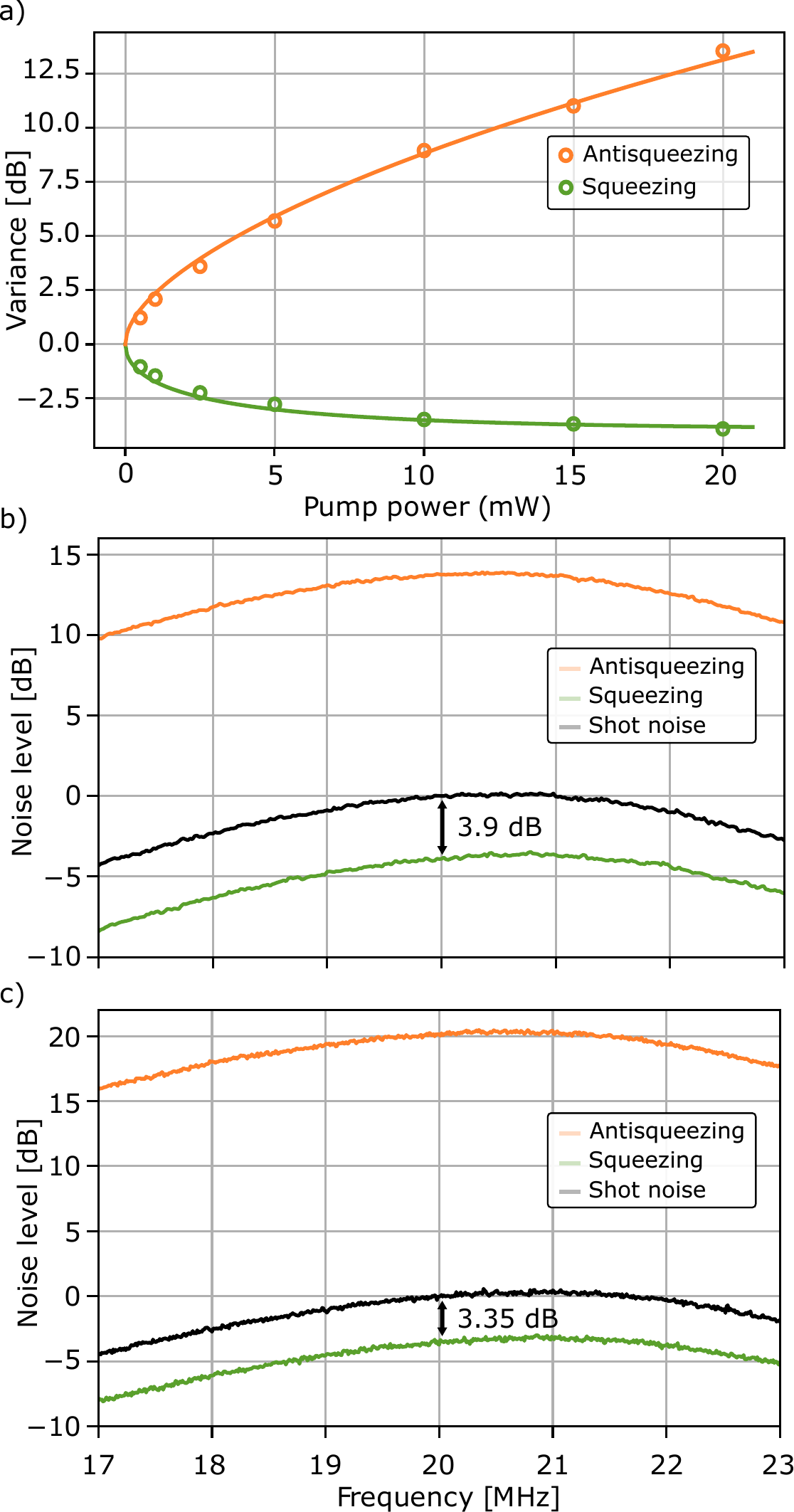} 
	\caption{\textbf{Squeezed light measurements} \textbf{a)} Squeezing (green) and antisqueezing (orange) detected in homodyne detection, as a function of pump power. Circles indicate experimental data, solid lines show fits to Eq.(1). \textbf{b)} Squeezing (green) and antisqueezing (orange) spectra measured in homodyne detection around $20~\mathrm{MHz}$ for $20~\mathrm{mW}$ of pump power. The black trace is the shotnoise level. A squeezing level of $-3.9~\mathrm{dB}$ is obtained after subtracting electronic noise $13.2~\mathrm{dB}$ below shot noise. \textbf{c)} Bright squeezing (green) and antisqueezing (orange) spectra with $3.2~\mathrm{mW}$ of coherent amplitude, measured in direct detection. Shot noise is the black trace, the pump power is 40 mW. A squeezing level of $-3.35~\mathrm{dB}$ is obtained after subtracting electronic noise $15.3~\mathrm{dB}$ below shot noise.} 
	\label{fig:fig3} 
\end{figure}

We vary the OPA pump power while monitoring the squeezing and antisqueezing levels at a sideband frequency of 20 MHz. 
Switching between the squeezed and antisqueezed quadratures is achieved by rotating the half-wave plate WP2. 
The shot-noise reference is obtained by blocking the OPA pump, thereby suppressing the signal at the PBS1 signal port.
Figure \ref{fig:fig3}(a) shows the squeezing and antisqueezing levels as a function of pump power after subtraction of the electronic noise (green and orange markers, respectively). Figure \ref{fig:fig3}(b) shows representative noise spectra of the squeezed, antisqueezed, shot-noise, and electronic-noise signals at a pump power of 20 mW. At this frequency, we observe $-3.61\pm0.05~\mathrm{dB}$ of squeezing and $13.54\pm0.02~\mathrm{dB}$ of antisqueezing. The electronic-noise clearance is $13.2~\mathrm{dB}$, corresponding to a squeezing level of $-3.9~\mathrm{dB}$ after electronic-noise subtraction.

To quantify the total detection efficiency and assess the impact of phase noise, we fit the measured squeezing and antisqueezing levels using a standard theoretical model that accounts for optical loss and phase-noise-induced leakage of antisqueezing into the squeezed quadrature:
\begin{equation}
    V^{\pm}(P) = \eta\left[e^{\pm2\alpha\sqrt{P}}cos^2\delta+ e^{\mp2\alpha\sqrt{P}}sin^2\delta \right] +1-\eta +EN
\label{eq:eq1}
\end{equation}

where $V^{-}$ and $V^{+}$ denote the variances of the squeezed and antisqueezed quadratures, respectively; $\eta$ is the total detection efficiency; $\alpha\sqrt{P}$ is the squeezing parameter with $P$ being the OPA pump power; $\delta$ represents phase noise; and EN denotes the electronic noise. $e^{\pm2\alpha\sqrt{P}}$ corresponds to the squeezing and antisqueezing levels generated in the waveguide. 
The experimental data are fit in dB units (solid green and orange lines in Fig. \ref{fig:fig3}(a)), yielding $\eta_{\mathrm{total,fit}} = 0.61 \pm 0.02$, $\delta = 12 \pm 30~\mathrm{mrad}$, and $\alpha = 12.4 \pm 0.1~\mathrm{mW^{1/2}}$. The large uncertainty in $\delta$ reflects the fact that the observed squeezing is primarily limited by optical losses rather than by phase noise.

The total efficiency $\eta_{\mathrm{total}}$ arises from several contributions: waveguide losses $\eta_{\mathrm{wg}}$ (including scattering, green-induced infrared absorption (GRIIRA), photorefractive effects and group velocity mismatch between the seed and pump \cite{bortz1995quasi,furukawa2001green,bazzan2015optical,jankowski2022quasi}), propagation losses between the waveguide and the detectors $\eta_{\mathrm{prop}} = 0.96$, spatial and temporal mode overlap between the LO and the squeezed field $\eta_{\mathrm{overlap}} = \eta_{\mathrm{spatial}}.\eta_{\mathrm{temp,exp}} = 0.97$, and detector QE $\eta_{\mathrm{det}} = 0.75$. The overall efficiency is therefore given by$    \eta_{\mathrm{total,exp}}=\eta_{\mathrm{wg}}.\eta_{\mathrm{prop}}.\eta_{\mathrm{overlap}}.\eta_{\mathrm{det}}$.

Comparing $\eta_{\mathrm{total,exp}}$ with the fitted value $\eta_{\mathrm{total,fit}}$ yields a waveguide efficiency of $\eta_{\mathrm{wg}} = 0.87$, consistent with the literature on LN waveguides \cite{eto2011efficient,bazzan2015optical}.
We estimate the squeezing generated in the waveguide $V^{-}_{wg}$ by correcting for all losses and propagating the uncertainties in $\eta_{\mathrm{total,fit}}$, $V^{-}$ and $\delta$ in Eq.(1). This yields $V^{-}_{wg}=-15.4\!\mrel{+ 2.7\\[.12ex]-8.7} \mathrm{dB}$. The large lower uncertainty limit arises from operation near the loss-limited lower bound of the measured squeezing. Practical SRS implementations remain subject to waveguide and displacement losses, limiting the bright squeezing level to $-6.2~\mathrm{dB}$. For other on-chip squeezing applications, where only waveguide losses are considered, the corresponding squeezing level is $-8.1~\mathrm{dB}$.

We also perform direct measurement of the bright squeezing, corresponding to the detection scheme used in nonlinear microscopy where the signal is encoded in the optical intensity.
To this end, the LO power is increased to approximately $31~\mathrm{mW}$ after the waveguide, and the splitting ratio at PBS1 is set to 90/10, resulting in $3.2~\mathrm{mW}$ directly measured on the photodetector. 
The reduced fraction of the squeezed field reaching the detector introduces an additional effective loss factor of $\eta_{\mathrm{direct}}=0.9$. 
Figure \ref{fig:fig3}(c) shows the measured squeezing and antisqueezing levels relative to the shot-noise reference for the bright squeezed beam. A pump power of 40 mW was used, accounting for the larger level of antisqueezing. The measured squeezing level was $-3.2~\mathrm{dB}$, and $-3.35~\mathrm{dB}$ after subtracting the electronic noise. This value is consistent with the squeezing measured previously in homodyne detection: $-3.9~\mathrm{dB}$ decreased by an additional 10\% of shot noise yields $-3.3~\mathrm{dB}$ of observed squeezing.

\section{Conclusion}\label{secOPA}

In conclusion, we demonstrate a simple experimental setup that achieves high levels of bright amplitude ps pulsed squeezing. The approach relies on co-propagation of the LO and squeezed field through a single-mode waveguide to achieve high spatial overlap, with waveplates used to control their relative phase and squeezing displacement. We verify spatial and temporal mode overlaps of 99.7\% and 97.7\%, respectively. We measure $-3.35~\mathrm{dB}$ of bright squeezing in a direct-detection configuration at a coherent power of $3.2~\mathrm{mW}$, compatible with state-of-the-art nonlinear microscopy. After correcting for losses, we infer $-15.4^{+2.7}_{-8.7}~\mathrm{dB}$ of squeezing generated within the waveguide. The measured squeezing is primarily limited by optical losses, in particular the photodetector QE. High QE photodiodes are available and can be incorporated and after correcting for propagation and detection losses, the expected bright squeezing level for SRS microscopy is $-6.2~\mathrm{dB}$. By achieving this squeezing level at the photodamage limit of biological samples, an apparatus-independent quantum advantage in microscopy could soon be within reach.

\section*{Acknowledgements}
This research was supported by Defense Advanced Research Projects Agency (DARPA) INSPIRED
program (HR00112420356), the Air Force Office of Scientific Research under award numbers  FA9550-20-1-0391 and FA9550-22-1-0047, the Australian Research Council Centre of Excellence for Engineered Quantum Systems (EQUS, grant number CE170100009) and the Australian Research Council Centre of Excellence in Quantum Biotechnology (QUBIC,grant number CE230100021). J. Q. Grim was supported by the U.S. Office of Naval Research.

\section*{Competing Interests}
The other authors declare no competing interests.





\bibliography{sn-bibliography}

@article{zelmon1997infrared,
  title={Infrared corrected Sellmeier coefficients for congruently grown lithium niobate and 5 mol.\% magnesium oxide--doped lithium niobate},
  author={Zelmon, David E and Small, David L and Jundt, Dieter},
  journal={Journal of the Optical Society of America B},
  volume={14},
  number={12},
  pages={3319--3322},
  year={1997},
  publisher={Optical Society of America}
}

@article{schumaker1984noise,
  title={Noise in homodyne detection},
  author={Schumaker, Bonny L},
  journal={Optics letters},
  volume={9},
  number={5},
  pages={189--191},
  year={1984},
  publisher={Optical Society of America}
}

@article{terrasson2024fast,
  title={Fast biological imaging with quantum-enhanced Raman microscopy},
  author={Terrasson, Alex and Mauranyapin, Nicolas P and Casacio, Catxere A and Grim, Joel Q and Barnscheidt, Kai and Hage, Boris and Taylor, Michael A and Bowen, WP},
  journal={Optics Express},
  volume={32},
  number={21},
  pages={36193--36206},
  year={2024},
  publisher={Optica Publishing Group}
}

@Article{amari2023highly,
  author    = {Amari, Jorge and Takai, Junnosuke and Hirano, Takuya},
  journal   = {Optics Continuum},
  title     = {Highly efficient measurement of optical quadrature squeezing using a spatial light modulator controlled by machine learning},
  year      = {2023},
  number    = {4},
  pages     = {933--941},
  volume    = {2},
  publisher = {Optica Publishing Group},
  ranking   = {rank4},
}

@article{dong2008experimental,
  title={Experimental evidence for Raman-induced limits to efficient squeezing in optical fibers},
  author={Dong, Ruifang and Heersink, Joel and Corney, Joel F and Drummond, Peter D and Andersen, Ulrik L and Leuchs, Gerd},
  journal={Optics letters},
  volume={33},
  number={2},
  pages={116--118},
  year={2008},
  publisher={Optical Society of America}
}

@Article{eto2011efficient,
  author    = {Eto, Yujiro and Koshio, Akane and Ohshiro, Akito and Sakurai, Junichi and Horie, Keiko and Hirano, Takuya and Sasaki, Masahide},
  journal   = {Optics letters},
  title     = {Efficient homodyne measurement of picosecond squeezed pulses with pulse shaping technique},
  year      = {2011},
  number    = {23},
  pages     = {4653--4655},
  volume    = {36},
  publisher = {Optica Publishing Group},
  ranking   = {rank4},
}

@article{bortz1995quasi,
  title={Quasi-phase-matched optical parametric amplification and oscillation in periodically poled LiNbO3 waveguides},
  author={Bortz, ML and Arbore, MA and Fejer, MM},
  journal={Optics letters},
  volume={20},
  number={1},
  pages={49--51},
  year={1995},
  publisher={Optical Society of America}
}

@article{bazzan2015optical,
  title={Optical waveguides in lithium niobate: Recent developments and applications},
  author={Bazzan, Marco and Sada, Cinzia},
  journal={Applied physics reviews},
  volume={2},
  number={4},
  year={2015},
  publisher={AIP Publishing}
}

@article{jankowski2022quasi,
  title={Quasi-static optical parametric amplification},
  author={Jankowski, Marc and Jornod, Nayara and Langrock, Carsten and Desiatov, Boris and Marandi, Alireza and Lon{\v{c}}ar, Marko and Fejer, Martin M},
  journal={Optica},
  volume={9},
  number={3},
  pages={273--279},
  year={2022},
  publisher={Optical Society of America}
}

@article{furukawa2001green,
  title={Green-induced infrared absorption in MgO doped LiNbO 3},
  author={Furukawa, Y and Kitamura, K and Alexandrovski, A and Route, RK and Fejer, MM and Foulon, G},
  journal={Applied Physics Letters},
  volume={78},
  number={14},
  pages={1970--1972},
  year={2001},
  publisher={American Institute of Physics}
}

@Article{serkland1997amplitude,
  author    = {Serkland, DK and Kumar, Prem and Arbore, MA and Fejer, MM},
  journal   = {Optics letters},
  title     = {Amplitude squeezing by means of quasi-phase-matched second-harmonic generation in a lithium niobate waveguide},
  year      = {1997},
  number    = {19},
  pages     = {1497--1499},
  volume    = {22},
  publisher = {Optical Society of America},
}

@Article{Black2001,
  author  = {Eric D. Black},
  journal = {American Journal of Physics},
  title   = {An introduction to Pound–Drever–Hall laser frequency stabilization},
  year    = {2001},
}

@Article{kim1994quadrature,
  author     = {Kim, Chonghoon and Kumar, Prem},
  journal    = {Physical review letters},
  title      = {Quadrature-squeezed light detection using a self-generated matched local oscillator},
  year       = {1994},
  number     = {12},
  pages      = {1605},
  volume     = {73},
  publisher  = {APS},
  ranking    = {rank4},
  readstatus = {skimmed},
}

@article{xu2025pushing,
  title={Pushing the sensitivity of stimulated Raman scattering microscopy with quantum light: Current status and future challenges},
  author={Xu, Zicong and Nitanai, Sho and Oguchi, Kenichi and Ozeki, Yasuyuki},
  journal={Applied Physics Letters},
  volume={127},
  number={4},
  year={2025},
  publisher={AIP Publishing}
}

@article{heng2025quantum,
  title={Quantum-Enhanced Sensing with Squeezed Light: From Fundamentals to Applications},
  author={Heng, Xing and Zhang, Lingchen and Yin, Qingyun and Liu, Wei and Tang, Lulu and Zhai, Yueyang and Wei, Kai},
  journal={Applied Sciences},
  volume={15},
  number={18},
  pages={10179},
  year={2025},
  publisher={MDPI}
}

@Article{gong2023super,
  author    = {Gong, Li and Lin, Shulang and Huang, Zhiwei},
  journal   = {Optics Letters},
  title     = {Super-resolution stimulated Raman scattering microscopy enhanced by quantum light and deconvolution},
  year      = {2023},
  number    = {24},
  pages     = {6516--6519},
  volume    = {48},
  publisher = {Optica Publishing Group},
}

@article{li2024harnessing,
  title={Harnessing quantum light for microscopic biomechanical imaging of cells and tissues},
  author={Li, Tian and Cheburkanov, Vsevolod and Yakovlev, Vladislav V and Agarwal, Girish S and Scully, Marlan O},
  journal={Proceedings of the National Academy of Sciences},
  volume={121},
  number={45},
  pages={e2413938121},
  year={2024},
  publisher={National Academy of Sciences},
}

@article{davidovich1996sub,
  title={Sub-Poissonian processes in quantum optics},
  author={Davidovich, Luiz},
  journal={Reviews of Modern Physics},
  volume={68},
  number={1},
  pages={127},
  year={1996},
  publisher={APS}
}

@article{slusher1990quantum,
  title={Quantum Optics in the'80s},
  author={Slusher, RE},
  journal={Optics and Photonics News},
  volume={1},
  number={12},
  pages={27--30},
  year={1990},
  publisher={OSA}
}

@article{lawrie2020squeezing,
  title={Squeezing noise in microscopy with quantum light},
  author={Lawrie, Ben and Pooser, Raphael and Maksymovych, Petro},
  journal={Trends in Chemistry},
  volume={2},
  number={8},
  pages={683--686},
  year={2020},
  publisher={Elsevier}
}

@article{vahlbruch2016detection,
  title={Detection of 15 dB squeezed states of light and their application for the absolute calibration of photoelectric quantum efficiency},
  author={Vahlbruch, Henning and Mehmet, Moritz and Danzmann, Karsten and Schnabel, Roman},
  journal={Physical review letters},
  volume={117},
  number={11},
  pages={110801},
  year={2016},
  publisher={APS}
}

@article{casacio2021quantum,
  title={Quantum-enhanced nonlinear microscopy},
  author={Casacio, Catxere A and Madsen, Lars S and Terrasson, Alex and Waleed, Muhammad and Barnscheidt, Kai and Hage, Boris and Taylor, Michael A and Bowen, Warwick P},
  journal={Nature},
  volume={594},
  number={7862},
  pages={201--206},
  year={2021},
  publisher={Nature Publishing Group UK London}
}

@article{li2022quantum,
  title={Quantum-enhanced stimulated Brillouin scattering spectroscopy and imaging},
  author={Li, Tian and Li, Fu and Liu, Xinghua and Yakovlev, Vladislav V and Agarwal, Girish S},
  journal={Optica},
  volume={9},
  number={8},
  pages={959--964},
  year={2022},
  publisher={Optica Publishing Group}
}

@article{tan2023profiling,
  title={Profiling single cancer cell metabolism via high-content SRS imaging with chemical sparsity},
  author={Tan, Yuying and Lin, Haonan and Cheng, Ji-Xin},
  journal={Science Advances},
  volume={9},
  number={33},
  pages={eadg6061},
  year={2023},
  publisher={American Association for the Advancement of Science}
}

@article{cheng2021emerging,
  title={Emerging applications of stimulated Raman scattering microscopy in materials science},
  author={Cheng, Qian and Miao, Yupeng and Wild, Joseph and Min, Wei and Yang, Yuan},
  journal={Matter},
  volume={4},
  number={5},
  pages={1460--1483},
  year={2021},
  publisher={Elsevier}
}

@article{zhang2019spectral,
  title={Spectral tracing of deuterium for imaging glucose metabolism},
  author={Zhang, Luyuan and Shi, Lingyan and Shen, Yihui and Miao, Yupeng and Wei, Mian and Qian, Naixin and Liu, Yinong and Min, Wei},
  journal={Nature biomedical engineering},
  volume={3},
  number={5},
  pages={402--413},
  year={2019},
  publisher={Nature Publishing Group UK London}
}

@article{tian2016monitoring,
  title={Monitoring peripheral nerve degeneration in ALS by label-free stimulated Raman scattering imaging},
  author={Tian, Feng and Yang, Wenlong and Mordes, Daniel A and Wang, Jin-Yuan and Salameh, Johnny S and Mok, Joanie and Chew, Jeannie and Sharma, Aarti and Leno-Duran, Ester and Suzuki-Uematsu, Satomi and others},
  journal={Nature communications},
  volume={7},
  number={1},
  pages={13283},
  year={2016},
  publisher={Nature Publishing Group UK London}
}

@article{saar2010video,
  title={Video-rate molecular imaging in vivo with stimulated Raman scattering},
  author={Saar, Brian G and Freudiger, Christian W and Reichman, Jay and Stanley, C Michael and Holtom, Gary R and Xie, X Sunney},
  journal={science},
  volume={330},
  number={6009},
  pages={1368--1370},
  year={2010},
  publisher={American Association for the Advancement of Science}
}

@article{schiessl2019phenazine,
  title={Phenazine production promotes antibiotic tolerance and metabolic heterogeneity in Pseudomonas aeruginosa biofilms},
  author={Schiessl, Konstanze T and Hu, Fanghao and Jo, Jeanyoung and Nazia, Sakila Z and Wang, Bryan and Price-Whelan, Alexa and Min, Wei and Dietrich, Lars EP},
  journal={Nature communications},
  volume={10},
  number={1},
  pages={762},
  year={2019},
  publisher={Nature Publishing Group UK London}
}

@Article{xu2022quantum,
  author  = {Xu, Zicong and Oguchi, Kenichi and Taguchi, Yoshitaka and Takahashi, Shun and Sano, Yuki and Mizuguchi, Takaha and Katoh, Kazuhiro and Ozeki, Yasuyuki},
  journal = {Optics Letters},
  title   = {Quantum-enhanced stimulated Raman scattering microscopy in a high-power regime},
  year    = {2022},
  number  = {22},
  pages   = {5829},
  volume  = {47},
}

@Article{camp2015chemically,
  author  = {Camp Jr, Charles H and Cicerone, Marcus T},
  journal = {Nature photonics},
  title   = {Chemically sensitive bioimaging with coherent Raman scattering},
  year    = {2015},
}

@article{cheng2015vibrational,
  title={Vibrational spectroscopic imaging of living systems: An emerging platform for biology and medicine},
  author={Cheng, Ji-Xin and Xie, X Sunney},
  journal={Science},
  volume={350},
  number={6264},
  pages={aaa8870},
  year={2015},
  publisher={American Association for the Advancement of Science}
}

@Article{freudiger2008label,
  author    = {Freudiger, Christian W and Min, Wei and Saar, Brian G and Lu, Sijia and Holtom, Gary R and He, Chengwei and Tsai, Jason C and Kang, Jing X and Xie, X Sunney},
  journal   = {Science},
  title     = {Label-free biomedical imaging with high sensitivity by stimulated Raman scattering microscopy},
  year      = {2008},
  number    = {5909},
  pages     = {1857--1861},
  volume    = {322},
  publisher = {American Association for the Advancement of Science},
}

\end{document}